\documentclass{PoS}

\title{Time variability of low angular momentum accretion flows around black hole}

\ShortTitle{Time variability of low angular momentum accretion flows around black hole}

\author{\speaker{Ishika Palit}\\
        Center for Theoretical Physics, Polish Academy of Sciences, Al. Lotnikow 32/46, 02-668 Warsaw, Poland\\
        E-mail: \email{palit@cft.edu.pl}}

\author{Agnieszka Janiuk\\
        Center for Theoretical Physics, Polish Academy of Sciences, Al. Lotnikow 32/46, 02-668 Warsaw, Poland\\
        E-mail: \email{agnes@cft.edu.pl}}
        
\author{Petra Sukova\\
        Astronomical Institute, Czech Academy of Sciences, Bo\v{c}n\'{i} II 1401, CZ-141 00 Prague, Czech Republic\\
        E-mail: \email{petra.sukova@asu.cas.cz}}        
\abstract{We present relativistic 2D and 3D GRMHD simulation of  axisymmetric, inviscid, hydrodynamic accretion flows in a fixed Kerr black hole gravitational field. The flow is characterized by a low angular momentum with respect to a Keplerian one. A relativistic fluid where its bulk velocity is comparable to the speed of light, flowing in the accretion disk very close to the horizon should be described by the adiabatic index: $4/3 < \gamma < 5/3$. The time-dependent evolution of shock position and respective effect on mass accretion rate and oscillation frequency with  varying adiabatic index has been studied. Here we present some of the results for the adiabatic index = 1.4 in a 2D and 3D model.}

\FullConference{High Energy Phenomena in Relativistic Outflows VII - HEPRO VII\\
		9-12 July 2019\\
		Facultat de F\'{i}sica, Universitat de Barcelona, Spain}

\begin{document}

\section{Introduction}
Observations have shown that black hole accretion flows can produce high-energy X-ray or gamma-ray emission \cite{belloni2011black}. Such high-energy emission is likely to be produced close to the black-hole horizon. As the quality and quantity of high-energy observations improved over the years, it became evident that photons must be created in a hot, tenuous, advection dominated region called the corona. This corona, boiling violently above the comparatively cool disk, is very close to the event horizon of the black hole. The observed spectrum of black hole accretion flows is partly coming from a Keplerian disk in the form of multi-color black body emission \cite{shakura1973black} \cite{abramowicz1988slim} together with a power-law component that mainly comes from the quasi spherical, low angular momentum flow component. Sub-Keplerian accretion disk with possible multi-transonic flow has now been discussed in great details over the past few decades \cite{sandip1996} \cite{paczy1982}.
\\
It is well known that matter very far from the black hole having negligible speed, i.e., below that of the local sound speed, has to fall into the black hole with the speed of light to satisfy the causality condition. Thus the transonic mode of accretion is the most viable scenario as the flow must cross sonic points in order to turn from subsonic to supersonic \cite{sandip1996} \cite{paczy1982} \cite{fukue1987}. 
\\
The inviscid flow of infalling matter, rotating with low angular momentum, feels the centrifugal barrier near the horizon as the centrifugal force ($1/r^{3} $) increases rapidly compared to the gravitational force (which goes as $1/r^{2}$) as the distance r decreases. This initially slows down the matter typically through a shock transition, and matter piles up, eventually producing a possible density jump before entering the black hole supersonically.
\\
The general outcome of such inviscid accretion models in a general relativistic regime and with a relativistic equation of state (EoS), shows that the transonic, radiative, and thermodynamic behaviour of the flow are strongly related to its composition. The adiabatic index, $\gamma$ provides the information about microphysics of the flow and thus different values of $\gamma$ correspond to different types of astrophysical objects or different phases of astrophysical activity.
The temperature range for advective accretion flow is $10^{6}K < T < 10^{11}K$ which shows that the same value of $\gamma$ is not able to explain the whole scenario. In reality, adiabatic index must have a value in between 4/3 and 5/3. For example, weakly active galaxies, such as Sgr A, is usually studied using a polytropic EoS with $\gamma$ around $5/3$, because these flows are believed to be radiatively inefficient and gas pressure dominated. 
On the contrary, in the case of gamma-ray bursts (GRBs) the flow is thought to account for the presence of electron-positron pairs, nucleons, photons and neutrinos \cite{Janiuk2004}, and the accretion flow is radiation pressure dominated, and requires a relativistic EoS with $\gamma = 4/3$. Also, even smaller values of the adiabatic index, close to the isothermal value of 1.0, may be relevant to systems like proto-planetary and proto-galactic disks.
\\
Among the parameters determining the sonic points and shock position in the transonic solution, the adiabatic index also seems to play a crucial role in determining the formation of standing or oscillating shocks \cite{sandip1996}. The oscillations of these shock fronts depend upon the physical properties of the matter and may give an explanation for the observed quasi periodic oscillations in black hole X-ray light curves. %However, most of the works to explain accretion models were done with fixed $\gamma$ (adiabatic index) equation of state (EoS). 
\\
In former works (see  e.g. \cite{sukova2017shocks}), the numerical studies of quasi-spherical transonic accretion and time-dependent evolution of the shock position were for the first time addressed in a fully general relativistic (GR) scheme. This is why the results obtained were not only qualitatively, but also quantitatively relevant for the observation of realistic sources, where the GR effects cannot be neglected in the closest vicinity of the black holes. These studies were however limited to the choice of a fixed parameter, namely the adiabatic index of $\gamma=4/3$. In a more recent work \cite{palit2019effects}, we explored other values of $\gamma$ to study the role of adiabatic index in producing different patterns of shock evolution. Here we present the follow-up of \cite{palit2019effects} by extending our simulation from 2D to 3D for a few models.

\section{Initial conditions}
To study the accretion flow of non-viscous matter, we start with the polytropic EoS: $ p = K \rho ^{\gamma}$
where $\gamma $ is the adiabatic index, p is the gas pressure and $\rho $ is the gas density. The details of the initial condition's prescription is given in  \cite{sukova2017shocks}, \cite{sukova2015shocks} and \cite{palit2019effects}.

Unlike for a thin disc, here  we assume the quasi-spherical distribution of the gas, provided by constant specific angular momentum $ \lambda $ \cite{abramowicz1981rotation}. Such a distribution of matter is possible to be formed instead of an evaporated Keplerian accretion disc. The flow transition from a supersonic to a subsonic region produced a shock at a given location, $(r_{s})$ . The region from $(r_{s})$  to the inner sonic point is defined as the post-shock region, which basically acts as a CENBOL (Centrifugal force dominated boundary Layer \cite{chakrabarti1989standing}.
The increase in density across the shock is described by the compression ratio, $\mathcal{R}_{comp} = \rho_{+}/\rho_{-}$  where $\rho_{+}$  the post-shock density and $\rho_{-}$ is the pre shock density. 
The initial conditions that we use here to study the $\gamma$ dependence of the sonic surface are similar to those used in previous studies (see e.g. \cite{sukova2015shocks}).
The rotation, i.e. the angular momentum of the flow, has been prescribed according to the relation 
\begin{equation}
  \lambda = \lambda^{eq} \sin^{2}\theta, 
  \label{eq:lambda}
\end{equation} 
where $\theta = \pi /2 $ and $\lambda^{eq}$ is the angular momentum in the equatorial plane.  The definition of angular momentum goes as usual,  $\lambda = \frac{u^{\phi}}{u^{t}}$. where $u^{\phi}$ and $u^{t}$ are radial velocities in $\phi$ and t direction. 
Our initial state does not correspond to the stationary state, because that is derived for a quasi-spherical distribution of gas with constant angular momentum in a pseudo-Newtonian potential. However, as we scale angular momentum according to (\ref{eq:lambda}) and we are in the general relativistic regime, the resulting configuration is not the solution of the stationary time-independent equations. Hence, it is expected that at the beginning of the simulation during a transient time, the flow adjusts itself into the appropriate profile.

\subsection{Transonic accretion}
A transonic flow of matter consists of three real critical points. The initial subsonic flow enters through the outer boundary and becomes supersonic at the outer sonic point, which in our case coincides with the outer critical point. Then it may jump from the supersonic regime to the subsonic one at ($r_{s}$) \cite{chakrabarti1989standing}.  
The region from $r_{s}$ to the inner critical point is defined as the post-shock region, which basically acts as a CENBOL region.
We get two values for the velocity gradient at the critical point, which are obtained as real and opposite to each other. This shows that the nature of the critical points are of saddle type  \cite{palit2019effects}.
The models are parameterized by the value of the specific angular momentum $\lambda $, the polytropic exponent $\gamma$ and the energy $\epsilon$, which set the critical point's position $r_{c}^{in}$, $r_{s}$ $r_{c}^{out}$ \cite{chakrabarti1996accretion}.

\subsection{Time evolution}
The model in 2D has been evolved up to t$_{final} = 10^{6}[M]$ whereas the final time of evolution in our 3D model is t$_{final} = 10^{4}[M]$. The units in the numerical simulation for the models are in geometrical units $(G = c = 1, [r] = [t] = [\lambda] =[M])$. The inner mass accretion rate has been calculated as in \cite{gammie2004black}:
\begin{equation}
    \dot{M}(r,t) = \int \rho u^{r} \sqrt{-g} d\theta d\phi
\end{equation}

\section{Numerical setup}
%\label{sec : time evolution}
The quasi spherical, slightly rotating flow in all the models in our simulations starts with the initial condition prescribing the critical point  and the velocity gradient at the critical point derived in \cite{sukova2015shocks}. Values for the model parameters as the specific energy ($\epsilon$), the specific angular momentum ($\lambda$), the adiabatic index ($\gamma$), the spin of black hole (a) and the distribution of angular momentum are set as the initial condition, which set the properties of the flow. The evolution of a non-magnetized gas (as assumed in our initial conditions) is simulated with the HARM package supplied with a few modifications (see  \cite{sukova2017shocks} and  \cite{palit2019effects}).
HARM (high-accuracy relativistic magnetohydrodynamics), is a conservative, shock-capturing scheme for evolving the equations of general relativistic MHD \cite{gammie2003harm}.

\subsection{Grid setup} 
The inner and outer radius of the computational grid  R$_{in}$ and R$_{out}$ are set as 0.2[M] and 50000[M]. The resolution for thte 2D and 3D models have been chosen as $[384*256*1]$ and $[384*256*92]$ in radial, polar and azimuthal directions respectively.

\section{Results}
\begin{figure}
\includegraphics[width = 0.5\textwidth]{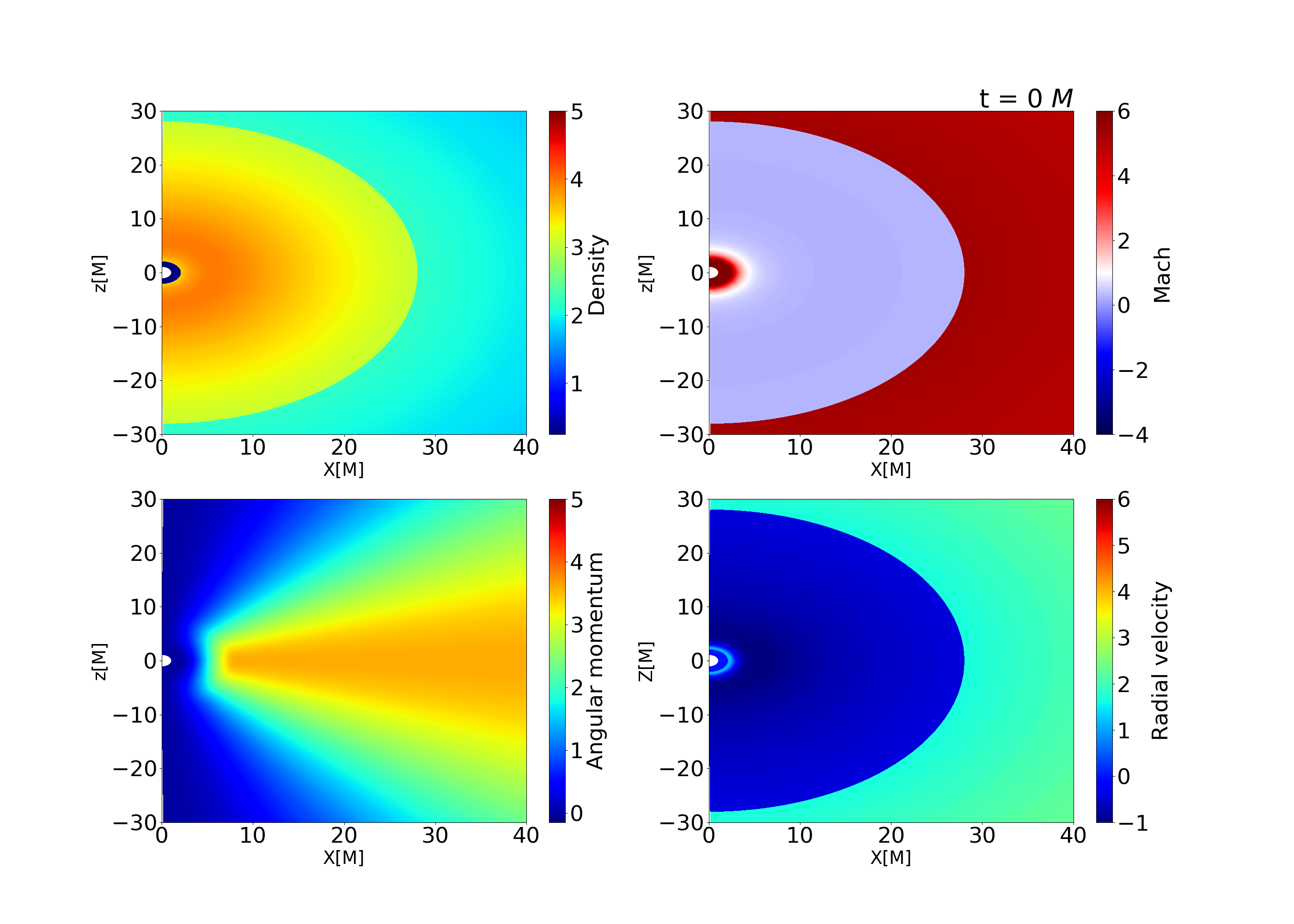}
\includegraphics[width = 0.5\textwidth]{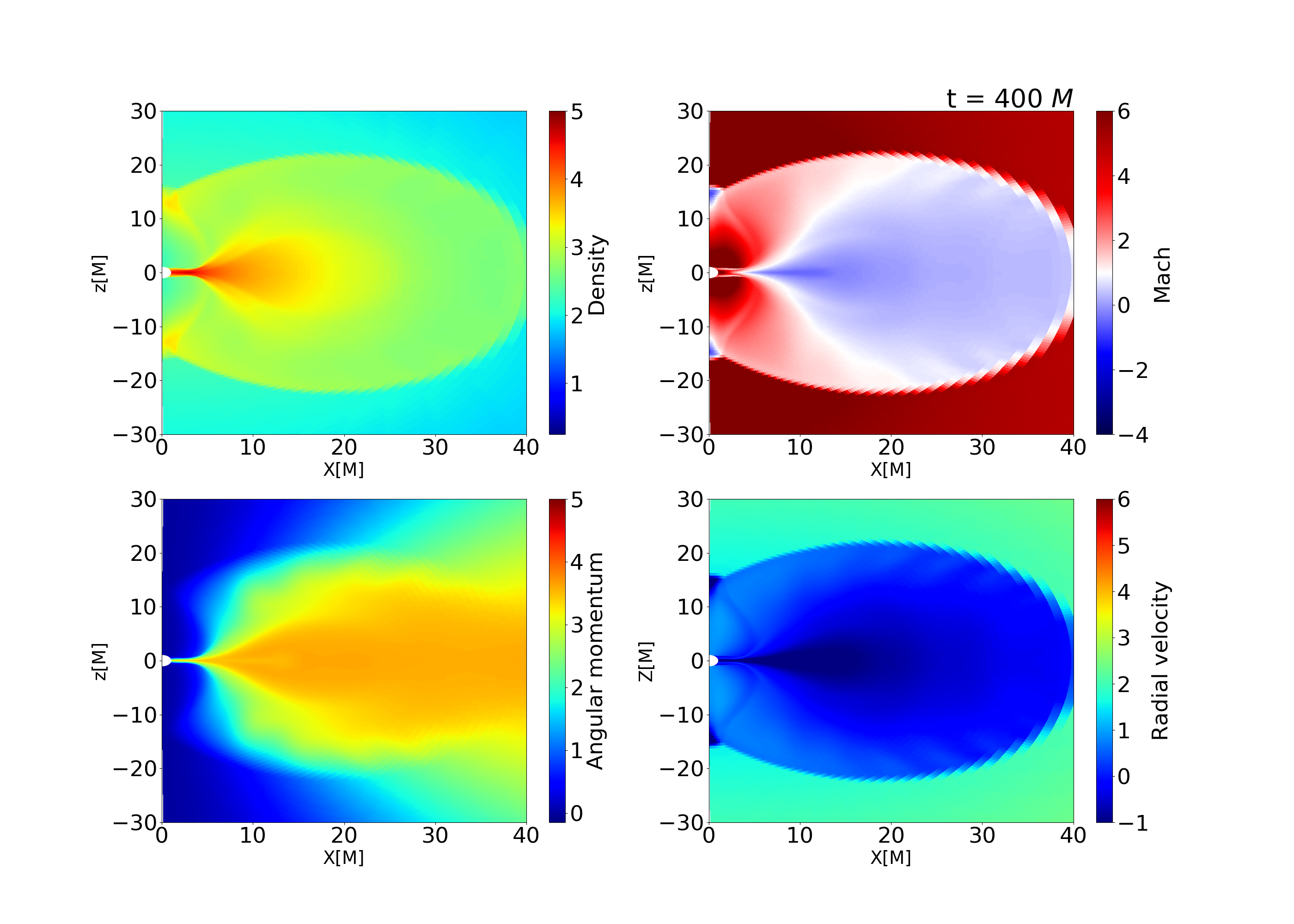}
\includegraphics[width = 0.5\textwidth]{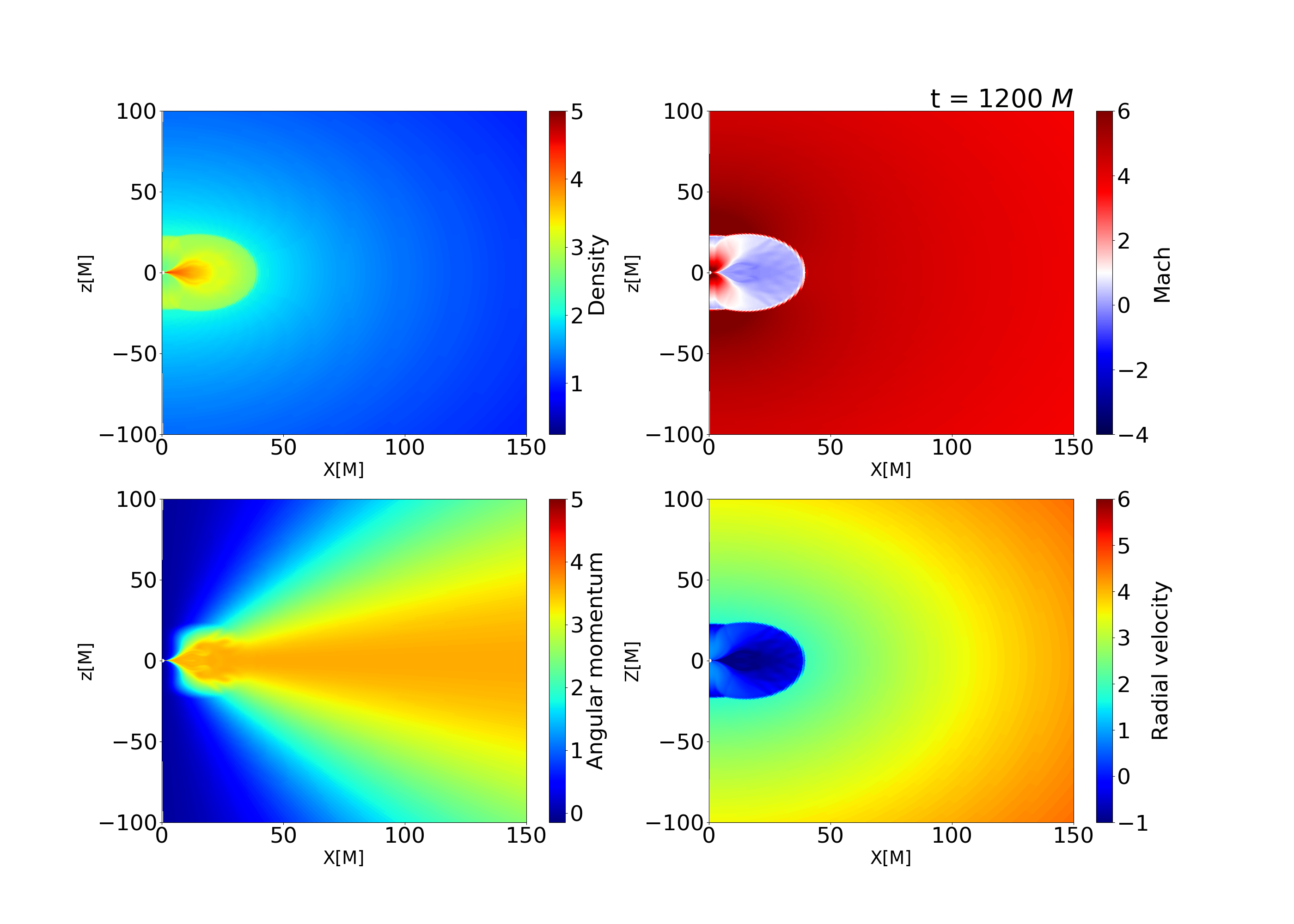}
\includegraphics[width = 0.5\textwidth]{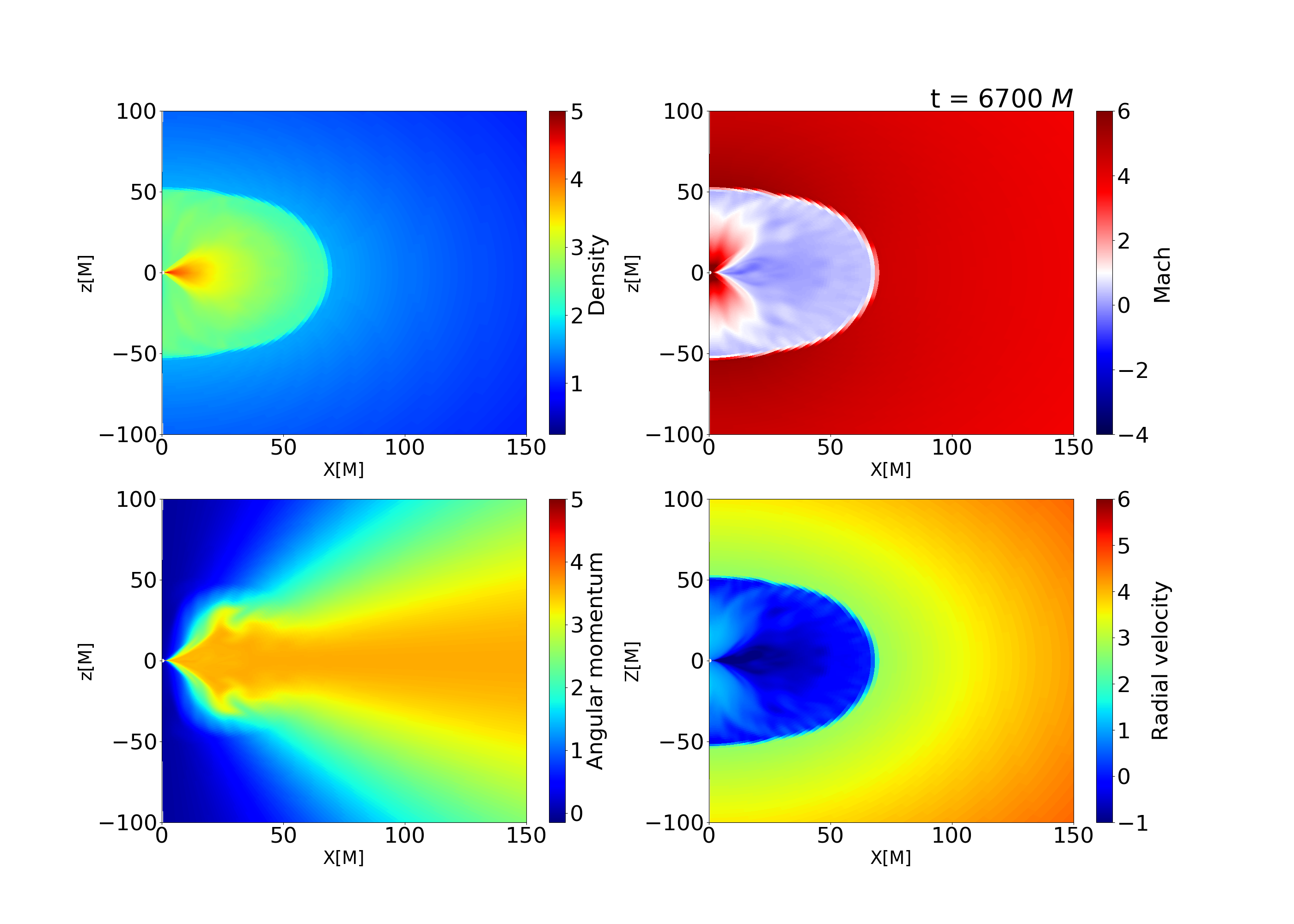}
 %\begin{center}
 % \centering
 %\includegraphics[width = 0.6\textwidth]{./figures/xz_2_2.png} 
% \end{center}
\caption{Density, Mach number, angular velocity and radial velocity profile for a model with $[\gamma = 1.4, \lambda =3.6  \rm [M], \epsilon = 0.0001 ]$ from 3D model G6 at four different time snapshots. }
\label{fig1}
\end{figure}

\begin{figure}
\textbf{t = 0[M]}\\
 \includegraphics[width =\textwidth]{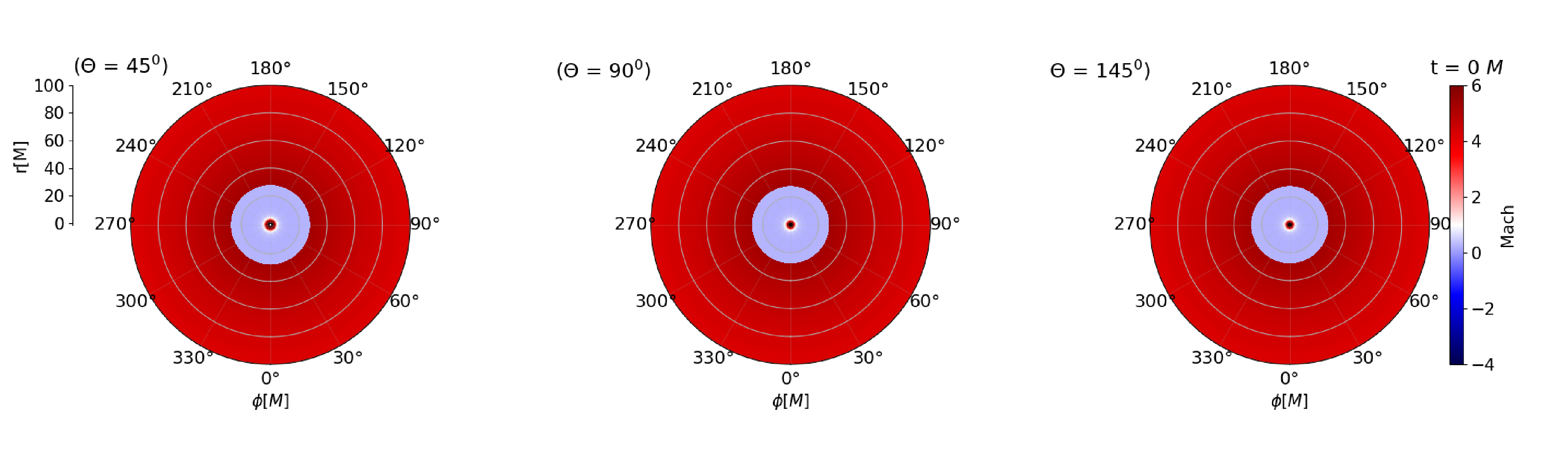}\\
% \textbf{t = 498000[M]}\\
 %\includegraphics[width = 1.0\textwidth]{./figures/img_1.png}\\
 \textbf{t = 6700[M]}\\
 \includegraphics[width =\textwidth]{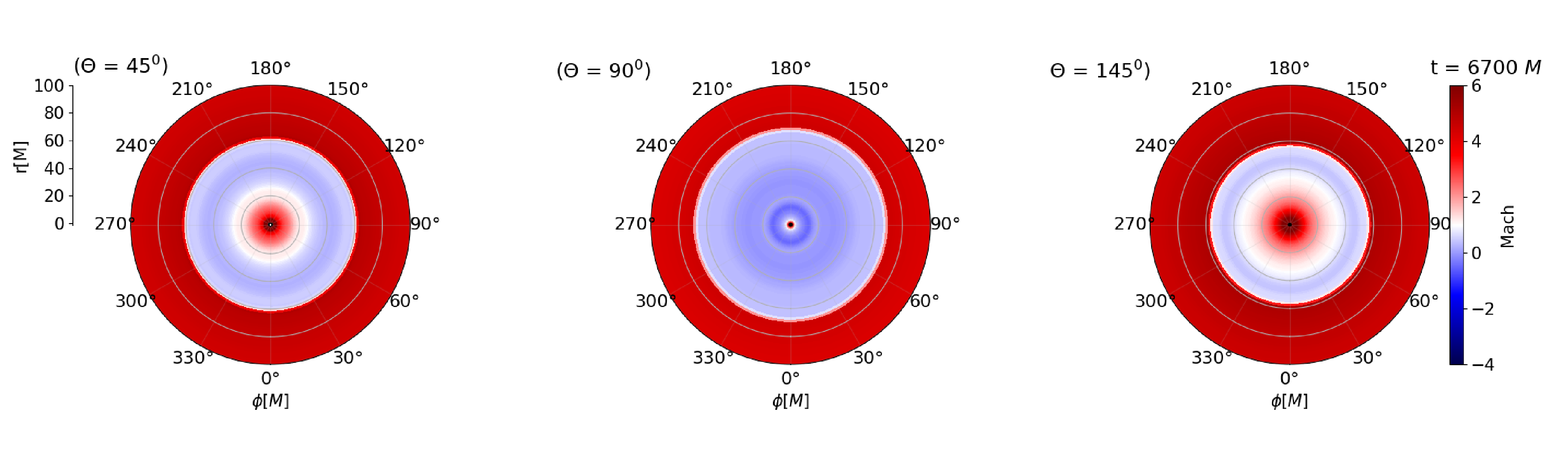}
 \caption{Mach number profile for model G6 with $[\gamma = 1.4, \lambda =3.6  \rm [M], \epsilon = 0.0001 ]$  at two different time snapshots. }
\label{fig2}
\end{figure}

The different behaviour of transonic flows corresponding to different parameter sets have been presented and discussed in extension in \cite{palit2019effects}. Here in addition to these results, we present two new models G6 and I4.
Model G6 is a 3D model similar to the 2D model D6 in \cite{palit2019effects} with parameters $[\gamma = 1.4, \lambda =3.6  \rm [M], \epsilon = 0.0001 ]$ from \cite{palit2019effects}. In model G6, the flow has been followed in phi direction as well, accounting for 92 grid points.
3D simulations are more realistic than 2D simulations but due to the axially symmetric nature of the Kerr metric, the results in 2D and 3D models are similar to each other.
Model I4 is a 2D model with similar parameters as the 2D model H4 $[\gamma = 1.4, \lambda =3.6  \rm [M], \epsilon = 0.0001 , a = 0.1]$ from \cite{palit2019effects} but with higher spin $(a = 0.8)$.
\\
Figure \ref{fig1} shows the cartesian plots for the 3D model G6, where the spherical Boyer Linquidst coordinates $(r, \theta, \phi)$ have been transformed to X,Y, Z as :
\begin{equation}
X = r * sin (\theta) *cos (\phi)
\end{equation}
\begin{equation}
Y = r * sin (\theta) *sin (\phi)
\end{equation}
\begin{equation}
Z = r * cos (\theta)
\end{equation}
The figure panel consist of of four quantities plotted in X and Z direction: the density, the Mach number, the angular momentum, and the radial velocity in clockwise direction.
\\
The Mach number is a dimensionless quantity defining the ratio of local flow velocity w.r.t local sound speed. When the radial flow velocity is larger than the sound speed, it travels supersonically (red-color in Fig.\ref{fig2}). Otherwise the flow is subsonic (blue-color in Fig.\ref{fig2}). The growth of a shock-driven bubble can be seen in different time slices. It has been observed in 2D model D6 that such shock-driven bubble displayed both vertical and horizontal oscillations along the flow time evolution (see \cite{palit2019effects}). The oscillation of the shock bubble in 2D model D6 was observed around t $\sim 10^{5}[M]$ which is longer time than t$_{final}$ for 3D model G6. Thus in model G6 only the growth of the shock bubble has been followed.
In Figure \ref{fig1}, the scale in X and Z axis has been enlarged in time snapshots placed in the second row to show the expansion of of the shock-drive bubble.

\begin{figure}
\textbf{t = 0[M]}\\
 \includegraphics[width =\textwidth]{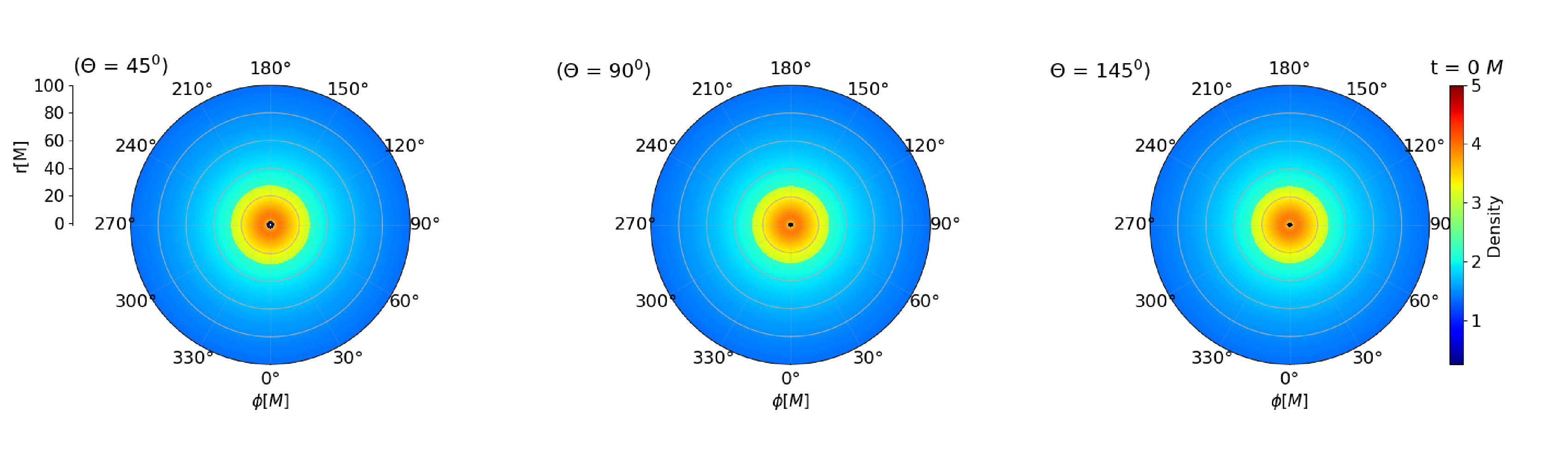}\\
% \textbf{t = 498000[M]}\\
% \includegraphics[width = 1.0\textwidth]{./figures/mach_15.png}\\
 \textbf{t = 6700[M]}\\
 \includegraphics[width =\textwidth]{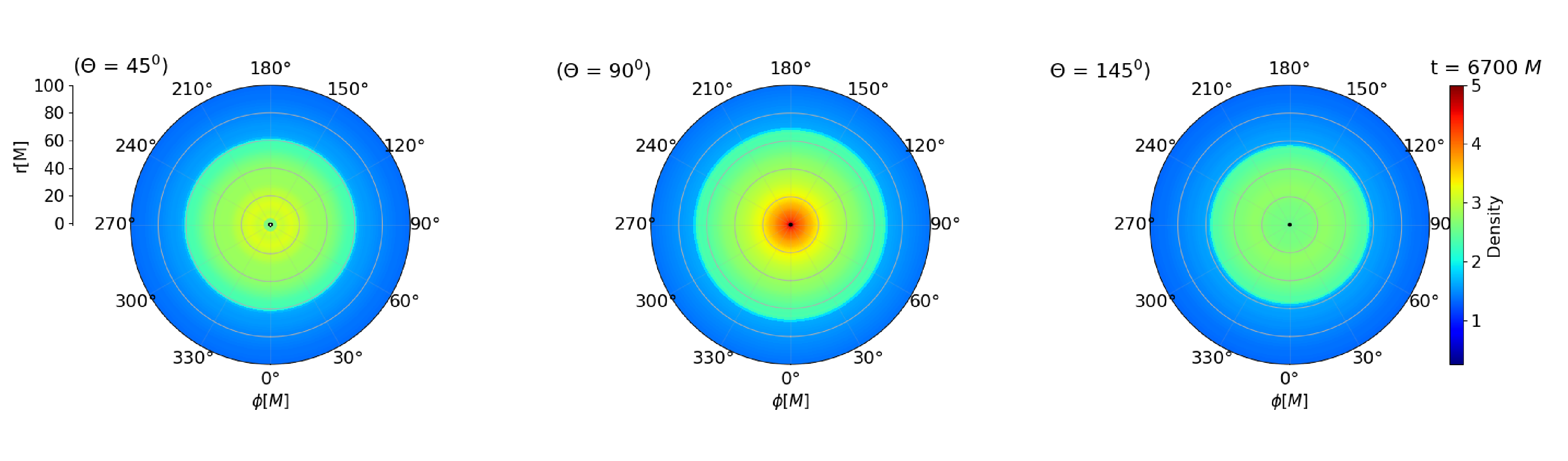}
\caption{Density profile for model G6 with $[\gamma = 1.4, \lambda =3.6  \rm [M], \epsilon = 0.0001 ]$ from 3D simulations at two different time snapshots.}
\label{fig3}
\end{figure}

Figure \ref{fig2} and \ref{fig3} are the polar plots from the similar 3D model G6. These are depicting Mach number and density of the flow respectively from three different angles at two different times.
The Mach distribution in the second panel in Figure \ref{fig2} is completely subsonic near the black hole at  $\theta = 90^{o}$ as it is not possible to distinguish  the supersonic inflow from the polar sides on the equator slice. When viewed from slant angles, such as $\theta = 45^{o}$ and  $\theta = 145^{o}$, the supersonic flow through the corners of the shock bubble can be identified. For longer runs, the  oscillations in the radial direction are expected to be seen more clearly in r-$\phi$ plots as the size of the subsonic part will vary if seen from $\theta = 45^{o}$ and  $\theta = 145^{o}$ at the same time. The density distribution in polar plots in Figure \ref{fig3} can be seen to display a maximum for $\theta = 90^{o}$ , i.e. at the equator, and with an almost identical profile for  $\theta = 45^{o}$ and  $\theta = 145^{o}$, similar to the results obtained from 2D simulations. We note that allowing for a longer computational time more accurate locations for the oscillating shock-driven bubble may be obtained. This, in turn, may allow to better estimate the oscillation frequencies, which can then be compared to the observed quasi-periodic oscillations in accretion flows (QPOs)

\begin{figure}
\includegraphics[width = 0.5\textwidth]{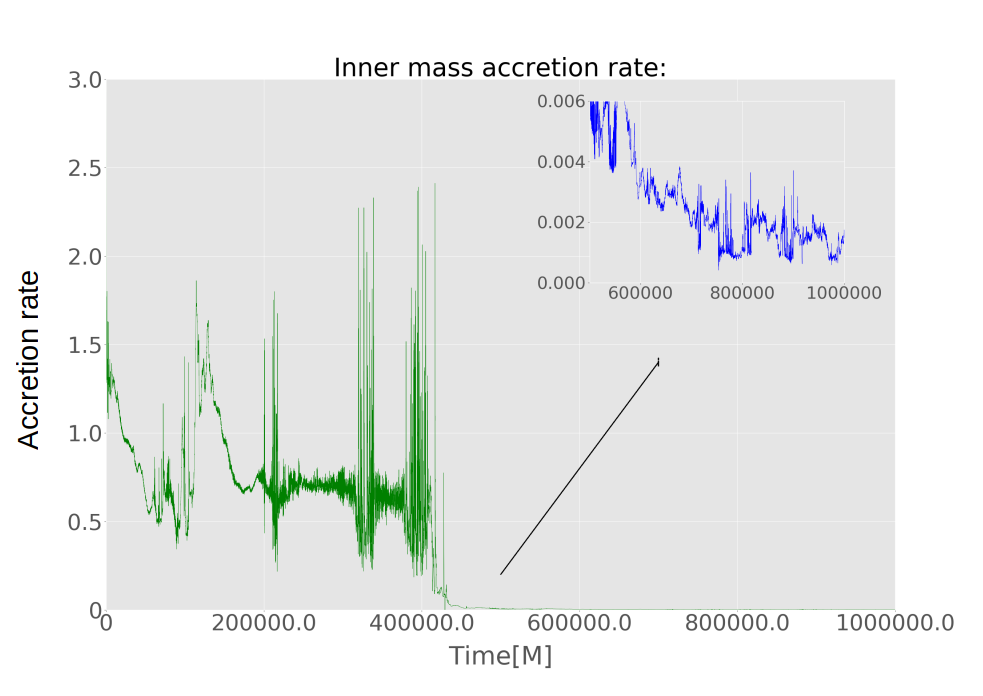}
\includegraphics[width = 0.5\textwidth]{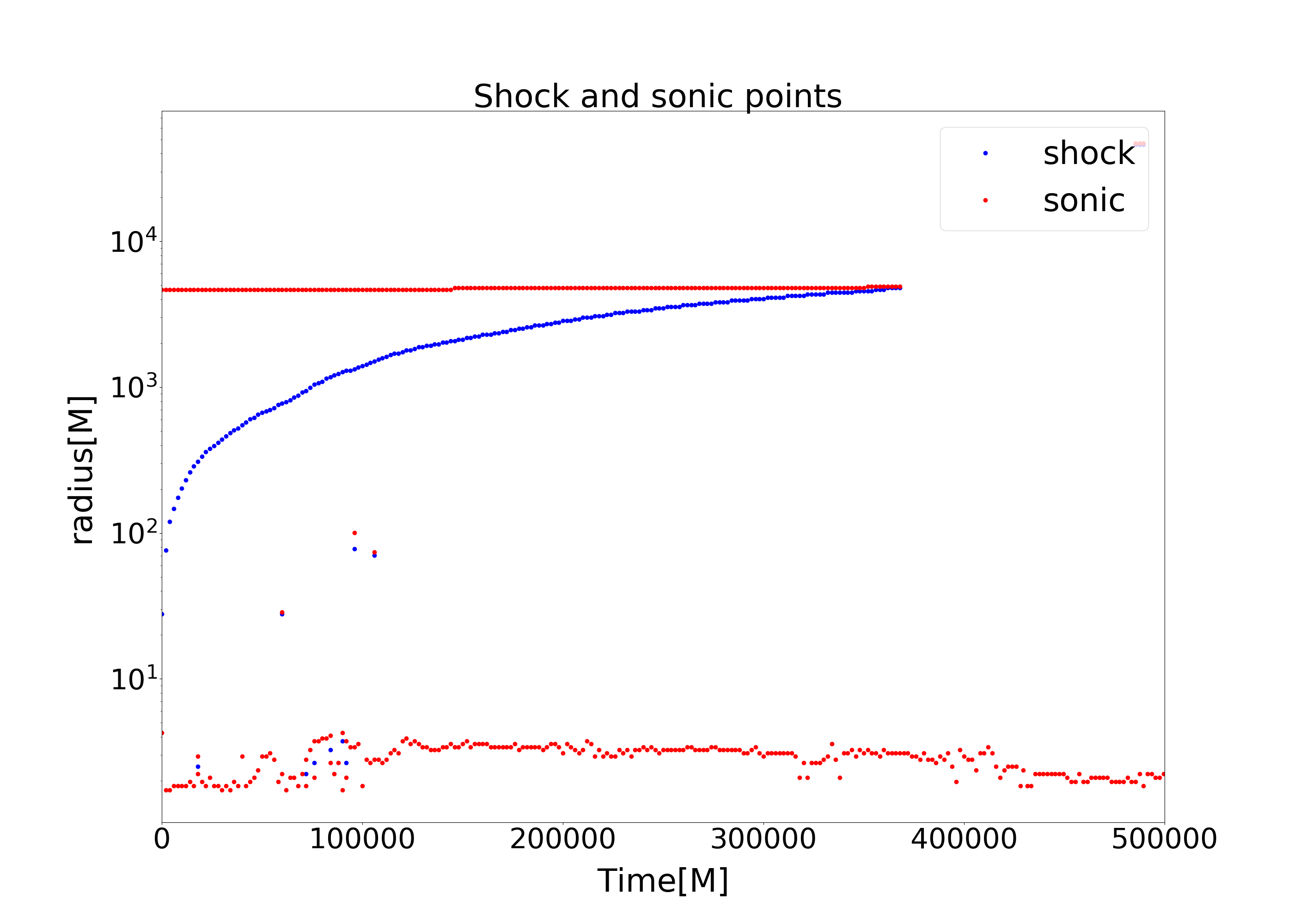}
\caption{Left panel shows the mass accretion rate for model I4 with $[\gamma = 1.4, \lambda =3.6 , \epsilon = 0.0001, a = 0.8 ]$ and right panel shows the evolution of shock position with time.}
\label{fig4}
\end{figure}

Figure \ref{fig4} shows the mass accretion rate and the shock evolution with time for a model with  $[\gamma = 1.4, \lambda =3.6  \rm [M], \epsilon = 0.0001 , spin(a) = 0.8]$ from 2D simulations. This is similar to our model H4 from  \cite{palit2019effects} but with higher values for the spin parameter (= 0.8 in this case). Here we can see that our shock bubble expands with time through the outer sonic point unlike in our previous model where the shock bubble had significant oscillations for small values of the black-hole spin. This result is consistent with \cite{sukova2017shocks}  where it has been shown that with higher spin it is required to have lower angular momentum of the flow so that the shock can oscillate. The mass accretion rate has been zoomed out in the inset in panel 1 of Figure \ref{fig4} to show the accretion rate from only the inner sonic point  after the shock gets expanded through the outer sonic point.

\section{Astrophysical significance}
Our results are important in connection to the observations of low frequency QPO's (LFQPO) in microquasars, where QPOs have been observed in the range from a few hundred mHz up to a few tens of Hz.
In particular, LFQPOs habe been observed in the frequency range of 0.05Hz - 10Hz in many x-ray binaries and microquasars such as GRO J1655-40 , XTE J1118-480, XTE J1748-288, IGR J17091-3624 (\cite{revnivtsev2000discovery}, \cite{vignarca2003tracing}, \cite{remillard1999rxte} ,\cite{iyer2015determination}, \cite{altamirano2012low}).
We found that several models from our simulations show subtle and significant shock front oscillations over time. The value for the oscillation frequency obtained in our simulations is representative of the time variability found in the above sources. We get an estimated oscillation
frequency range between 0.3Hz and 0.6Hz (from model D6), between 0.13Hz and 0.24Hz (from model H4), and between 4.8Hz and 8.6Hz (from model D3) (see \cite{palit2019effects} for details of these models).
\\
Our frequency range fits broadly into the observed frequencies of these systems. %Our simulation results are also in good agreement with the frequency range (0.1Hz - 15Hz) observed in black hole system GRO J1655-40.
%as well as the estimated accretion rate. 
Our model is therefore able to explain the LFQPOs observed in some microquasars, under the assumption that these oscillations are produced in the inner parts of an accreting adiabatic flow described with $\gamma = 1.4$ onto both a spinning and a non-spinning black-hole Further investigation is needed to study in detail the range of spins of the Kerr black holes that
is able to produce an oscillatory shock behaviour for low
angular momentum accretion flows. 
  Finally, our investigations of the shock front oscillation may also be relevant in the case of active galactic nuclei (AGN). Here the observations are not as precise as in the case of Galactic X-ray binaries, but the combined constraints from the energy spectrum and variability show that the soft excess is likely arising from the low-temperature Comptonization of the disc. This remains more or less constant on short time-scales, diluting the QPOs and the rapid variability, as it is seen e.g. in the Seyfert galaxy RE J1034 \cite{Middleton2009}.
  Also, after careful modeling of the noise continuum, a $\sim 3.8$ hr QPO
  was found in the ultrasoft AGN candidate 2XMM J123103.2+110648 \cite{Lin2013}. Such tentative detection might suggest that the shock front in this AGN oscillates in several modes (equatorial, polar, azimuthal), as suggested by our results.  
    
 \section{Acknowledgements}
 We thank Konstantinos Sapountzis and Ireneusz Janiuk for helpful discussions and computational support.This work was partially supported by the grant No. DEC-2016/23/B/ST9/03114 from the Polish National Science Center. We also acknowledge support from the Interdisciplinary Center for Mathematical Modeling of the Warsaw University, through the computational grant Gb79-9, and the PL-Grid computational resources through the grant grb2. PS is supported from Grant No. GACR-17-06962Y from Czech Science Foundation.


\begin{thebibliography}{99}


\bibitem{belloni2011black}
Belloni, T. M. and Motta, S. E. and Mu{\~n}oz-Darias, T, 'Black hole transients', MNRAS, Volume 410, 679-684, 2011 

\bibitem{shakura1973black}
Shakura, N. I. and Sunyaev, R., 'Black holes in binary systems: Observational appearance', A$\&$A, 24, 337-355, 1973

\bibitem{abramowicz1988slim}
Abramowicz, MA and Czerny, B and Lasota, JP and Szuszkiewicz, 'Slim accretion disks', ApJ, 332, 646-658, 1988

\bibitem{sandip1996}
Molteni, Diego and Ryu, Dongsu and Chakrabarti, Sandip K., 'Numerical Simulations of Standing Shocks in Accretion Flows around Black Holes: A Comparative Study', ApJ, 10.1086/177877, 1996

\bibitem{paczy1982}
Muchotrzeb, B. and Paczynski, B.,'Transonic accretion flow in a thin disk around a black hole', ACTAA, 32, 1-2, 1982

\bibitem{fukue1987}
Fukue, Jun, 'Transonic disk accretion revisited',Astronomical Society of Japan, Publications (ISSN 0004-6264), vol. 39, no.2, 1987

\bibitem{Janiuk2004}
Janiuk A., Perna R., Di Matteo T., Czerny B.,'Evolution of a neutrino-cooled disc in gamma-ray bursts', MNRAS, 355, 950, 2004

\bibitem{sukova2017shocks}
Sukov{\'a}, P and Charzy{\'n}ski, S and Janiuk, A, 'Shocks in the relativistic transonic accretion with low angular momentum',  MNRAS , 472, 4, 4327-4342, 2017

\bibitem{molteni1996numerical}
Molteni, D. and Ryu, D. and Chakrabarti, S. K,'Numerical simulations of standing shocks in accretion flows around black holes: A comparative study', astro-ph/9605116, 1996

\bibitem{das1999mass}
Das, T. and Chakrabarti, S. K, 'Mass outflow rate from accretion discs around compact objects', Classical and Quantum Gravity, IOP publishing, 1999


\bibitem{abramowicz1981rotation}
Abramowicz, MA and Zurek, WH, 'Rotation-induced bistability of transonic accretion onto a black hole', APJ, 246, 314-320, 1981

\bibitem{chakrabarti1989standing}
Chakrabarti, S. K, 'Standing Rankine-Hugoniot shocks in the hybrid model flows of the black hole accretion and winds', ApJ, 347, 365-372, 1989


\bibitem{sukova2015shocks}
Sukov{\'a}, P and Janiuk, A,'Shocks in the low angular momentum accretion flow', MNRAS, Volume 447, Issue 2, p.1565-1579, 2015

\bibitem{palit2019effects}
Palit, I and Janiuk, A and Sukova, P, 'Effects of adiabatic index on the sonic surface and time variability of low angular momentum accretion flows', MNRAS, 487, 1, 755--768, 2019

\bibitem{chakrabarti1996accretion}
Chakrabarti, S. K, 'Accretion processes on a black hole', Physics Reports,266, 5-6, 229--390, 1996

\bibitem{gammie2004black}
Gammie, C. F and Shapiro, S. L and McKinney, J. C, 'Black hole spin evolution', ApJ, 602, 1, 312, 2004

\bibitem{gammie2003harm}
Gammie, C. F and McKinney, J. C and T{\'o}th, G, 'HARM: a numerical scheme for general relativistic magnetohydrodynamics', ApJ, 589, 1, 444, 2003


\bibitem{revnivtsev2000discovery}
Revnivtsev, M and Sunyaev, R and Borozdin, K,'Discovery of 0.08 Hz QPO in the power spectrum of black hole candidate XTE J1118+ 480', Astronomy and Astrophysics, v.361, 2000

\bibitem{vignarca2003tracing}
Vignarca, Belloni, Psaltis, Dimitrios and Van Der Klis, 'Tracing the power-law component in the energy spectrum of black hole candidates as a function of the QPO frequency', A$\&$A, 2003
 
\bibitem{remillard1999rxte} 
Remillard, R. A and Morgan, E. H and McClintock, J. E and Bailyn, C. D and Orosz, J. A, 'RXTE Observations of 0.1-300 Hz Quasi-periodic Oscillationsin the Microquasar GRO J1655--40', ApJ, 522, 1, 397, 1999

\bibitem{iyer2015determination}
Iyer, N and Nandi, A and Mandal, S, 'Determination of the Mass of igr J17091--3624 From “spectro-temporal” Variations During the Onset phase of the 2011 Outburst', ApJ, 807, 1, 108, 2015

\bibitem{altamirano2012low}
Altamirano, Diego and Strohmayer, Tod, 'Low-frequency (11 mHz) oscillations in H1743-322: a new class of black hole quasi-periodic oscillations?', ApJ, 754, 2, L23, 2012


\bibitem{Middleton2009}
Middleton, M. and Done, C. and Ward, M., Gierli{\'n}ski, M. and 
	Schurch, N., 'RE J1034+396: the origin of the soft X-ray excess and quasi-periodic oscillation', MNRAS, 0807.4847, 394, 2009
	

\bibitem{Lin2013}
Lin, D. and Irwin, J.~A. and Webb, N.~A. and Barret, D. and Remillard, R.~A., 'Discovery of a Highly Variable Dipping Ultraluminous X-Ray Source in M94', ApJ, 1311.1198, 149, 2013





\end{thebibliography}
\end{document}